\documentclass[fleqn,usenatbib]{mnras}

\usepackage[T1]{fontenc}

\DeclareRobustCommand{\VAN}[3]{#2}
\let\VANthebibliography\thebibliography
\def\thebibliography{\DeclareRobustCommand{\VAN}[3]{##3}\VANthebibliography}

\usepackage{graphicx}	
\usepackage{amsmath}	
\usepackage{amssymb}	

\usepackage{amssymb}
\usepackage{longtable,ltxtable,booktabs}

\usepackage{multirow}
\usepackage{color}

\newcommand{\Msolar}{\mbox{\,$\rm {\rm M}_{\odot}$}} 

\def\approxge{\mathrel{\raise1.16pt\hbox{$>$}\kern-7.0pt \lower3.06pt\hbox{{$\scriptstyle\approx$}}}} 
\def\approxle{\mathrel{\raise1.16pt\hbox{$<$}\kern-7.0pt \lower3.06pt\hbox{{$\scriptstyle \approx$}}}} 

\title[The GW signals from merging NSWD binaries]
{The diverse morphology of gravitational wave signals from merging neutron-star white-dwarf binaries}

\author[]{Shenghua Yu$^{1}$\thanks{shenghuayu@bao.ac.cn}, Youjun Lu$^{2,3}$\thanks{luyj@nao.cas.cn}, C. Simon Jeffery$^{4}$\thanks{Simon.Jeffery@armagh.ac.uk}, 
Zhanwen Han$^{5}$\thanks{zhanwenhan@ynao.ac.cn}, DongDong Liu$^{5}$, Jie Yang$^{6}$,
\newauthor  
Xilong Fan$^{7}$, Bo Peng$^{1}$, Jianbin Li$^{1}$\\
$^1$State Key Laboratory of Radio Astronomy and Technology, National Astronomical Observatories, Chinese Academy of Sciences, \\ 20A Datun Road, Beijing 100101, China \\
$^2$CAS Key Laboratory for Computational Astrophysics, National Astronomical Observatories, Chinese Academy of Sciences, \\ 20A Datun Road, Beijing 100101, China \\
$^3$School of Astronomy and Space Sciences, University of Chinese Academy of Sciences, 19A Yuquan Road, Beijing 100049, China \\
$^4$Armagh Observatory and Planetarium, College Hill, Armagh BT61 9DG, N. Ireland\\
$^5$National Astronomical Observatories/Yunnan Observatory, Chinese Academy of Sciences, Kunming 650011, China \\
$^6$School of Physical Science and Technology, Lanzhou University, Lanzhou 730000, China \\
$^7$School of Physics and Technology, Wuhan University, Wuhan 430072, China
}

\date{Accepted XXX. Received YYY; in original form ZZZ}

\pubyear{2025}

\begin{document}
\label{firstpage}
\pagerange{\pageref{firstpage}--\pageref{lastpage}}
\maketitle

\begin{abstract}
In sufficiently compact neutron star-white dwarf (NSWD) binary systems, orbital decay means the white dwarf eventually fills its shrinking Roche lobe, initiating a phase of mass transfer. 
The exchange of angular momentum-both internal and external-plays a critical role in determining the binary's evolutionary outcome. 
For neutron stars with relatively low magnetic fields and spin frequencies, whether the orbital separation continues to shrink depends on the interplay between gravitational wave (GW) radiation and mass transfer dynamics. 
We compute the orbital evolution of NSWD binaries across a broad parameter space, incorporating four key variables. 
Our results reveal distinct boundaries in the NS-WD mass-mass diagram: binaries with white dwarf masses above these thresholds undergo rapid orbital decay and direct coalescence. 
The dependence of these boundaries on system parameters indicates that Roche-lobe-filling NSWD binaries can follow multiple evolutionary pathways -- a phenomenon we refer to as branched or polymorphic evolution.
NSWD binary systems emit strong and diverse GW signals, many of which would be detectable by space-based GW observatories. 
The morphology of the evolving GW waveform provides a direct diagnostic for the NSWD binary configuration, including any contribution from an accretion disk. 
Our models can provide critical waveform templates for identifying merging binary signals in real-time GW data. 
\end{abstract}


\section{Introduction}
\label{intro}

Interacting neutron star-white dwarf (NSWD) binaries are promising sources of gravitational waves (GWs) for space-based detectors. 
Some of these systems have extremely short orbital periods (10-80 minutes), emitting GWs in the frequency range of 0.001-0.8 Hz. 
This makes them particularly relevant for upcoming missions such as the Laser Interferometer Space Antenna (LISA) \citep{Nelemans10, Amaro-Seoane23}, Taiji \citep{Ruan20}, and TianQin \citep{Wang19}. 
Studying their orbital parameters (mass, period, and eccentricity), chemical composition, and spatial distribution can provide valuable insights into accretion processes under extreme conditions and the formation and evolution of compact binaries. 
These studies also inform our understanding of key physical processes such as common envelope evolution and mass transfer stability.

Theoretically, NSWD binaries are expected to be the second most numerous population of compact binaries in the Galaxy \citep{Amaro-Seoane23}. 
LISA is predicted to detect 100-300 such systems out of a total Galactic population of approximately $10^6-10^7$. Observationally, about 14 ultra-compact X-ray binaries (UCXBs)-five of which are in globular clusters-are believed to be low-mass NSWD systems with orbital periods between 600 and 3000 seconds. 
Among these, 4U 1728-34 and 4U 1820-30 have the shortest known periods, exhibiting coherent X-ray bursts at 646 s \citep{Galloway10} and 685 s \citep{Stella87}, respectively. 
The white dwarf masses in these systems typically range from 0.01 to 0.07 \Msolar.

Radio pulsar timing has identified approximately 162 pulsars in binary systems with likely white dwarf companions-131 in the Galactic disk and 31 in globular clusters 
(Australia Telescope National Facility pulsar catalogue, http://www.atnf.csiro.au/research/pulsar/psrcat, \citet{Manchester05} and references therein). 
Recent FAST GPPS survey has found another 83 pulsar binaries containing WD components in the Galaxy \citep{Wang25}, 
increasing the total candidate number of NSWDs to 245.  
Many of these companions have been optically confirmed through counterpart searches at the pulsars' astrometric positions. 
Spectroscopic observations allow measurements of the white dwarfs' effective temperatures, surface gravities, masses, and spectral types \citep{Edmonds01, Rivera-Sandoval15}.
The orbital periods of these pulsar-white dwarf binaries span from $\sim$0.037 days to 1230 days. 
The shortest-period system, discovered by FAST, has an orbital period of $\approx0.037$ days \citep{Pan23}. 
About eight short-period systems (four in the disk and four in globular clusters) are expected to merge within $10^{10}$ years due to GW emission.

Selection effects -- such as the intrinsic rarity of these systems, pulsar beaming, and signal smearing due to high orbital acceleration -- limit the detectability of tight pulsar binaries \citep{Ransom03}.
The lowest estimated white dwarf companion mass is $\approx0.07$ \Msolar, significantly below the typical minimum mass ($\approx0.17$ \Msolar) observed in double white dwarf systems via optical spectroscopy \citep{Kilic07, Brown13}. 
The highest reported mass, 1.58 \Msolar\ in PSR J1227-6208, slightly exceeds the Chandrasekhar limit, though this may be affected by the system's inclination angle \citep{Bates15}.

Previous studies have indicated that in NSWD binary systems, mass transfer from a WD onto a NS 
is dynamically unstable for WD masses greater than a critical value of $\sim0.5 M_{\odot}$ 
\citep{Verbunt88,Nelson86,Hut81}. \citet{Bobrick17} found that the critical mass of WD may be 
$\sim0.2 M_{\odot}$, binaries containing helium WDs with masses less than the critical value 
undergo stable mass transfer and evolve into ultracompact X-ray binaries. 
When the WD components in NSWD binaries have masses greater than the critical mass, it is usually 
considered that tidal disruption of the WD will take place \citep{Bobrick17,Bobrick22,Moran-Fraile2024}. 
A similar scenario was also applied to WD-black hole binaries \citep{Fryer99}. 

The mass loss or disruption of WD can form an accretion disk around the NS or black hole. 
The dynamical stability of a differentially rotating disk of fluid of uniform entropy and 
uniform specific angular momentum was investigated by \citet{Papaloizou84}. They found that 
the disk are unstable to low order non-axisymmetric modes and the modes grow on a dynamical time-scale. 
As a consequence of the instability, persistent orbiting matter clumps may rapidly grow, which 
could contribute detectable GW signals in the case of disks around spinning black holes \citep{Wessel2021}.
\citet{Moran-Fraile2024} simulated the merger of a 1.4$M_{\odot}$ NS with a 1$M_{\odot}$ Carbon-Oxygen
WD in the magnetohydrodynamic moving mesh code AREPO. They found that the frequency of GWs released during the 
merger is too high to be detectable by the LISA, but suitable for deci-hertz observatories such as LGWA, BBO, 
or DECIGO. 

The NSWD binary mergers may produce explosive transient events observable in electromagnetic radiation.  
\citet{Zenati19} employed hydrodynamical-thermonuclear 
FLASH-code simulations to study the evolution of WD debris discs formed following WD disruptions 
by NSs. They found that these explosions may be a new class of potentially observable 
transients. 
\citet{Zenati20} used similar simulations to model the disruption of CO-WDs by NSs, 
and found that the spectra of the explosive events share some similarities with rapidly evolving 
transients such as SN 2010X.  
\citet{Fernandez13b} studied the evolution of radiatively inefficient accretion disks 
generated by the merger of a WD and a neutron star or black hole. They found that 
detonations following a NSWD merger could account for some subluminous Type Ia supernovae, 
such as the class defined by SN 2002cx. 
In the high-energy band, \citet{Fryer99} and \citet{King2007b} considered the merger of 
a massive white dwarf with a neutron star as the source of gamma-ray bursts (GRBs), which 
would have a strong correlation with star formation, and occur close to the host galaxy. 

Extensive radio and X-ray observations suggest that NSWD binaries will contribute significantly to the GW signals detected by space-based observatories. 
Recently, gravitational waveform templates for various mass-transferring NSWD models have been developed \citep{Zhang24}, 
enabling matched-filter searches to extract binary evolution parameters from GW data \citep{Usman16}. 
However, the final evolutionary stages and corresponding waveforms of NSWD binaries remain poorly understood. 
Improved modelling of these processes will enhance the interpretation of GW signals and foster synergy between GW and electromagnetic observations.

In this study, we demonstrate that Roche-lobe-filling NSWD binaries can be classified into coalescing and non-coalescing systems. 
The mass threshold for coalescence depends on evolutionary parameters, indicating that environmental factors can lead to diverse evolutionary outcomes. 
We also explore the  GW signatures of NSWD binaries, informed by orbital parameters derived from radio observations. 
While low-mass NSWD binaries produce steady GW signals, high-mass coalescing systems may generate much stronger signals over shorter timescales.
Hence, given different environmental factors, an initial binary may evolve through one of several NSWD configurations, a property we call \textit{polymorphic evolution}, 
each configuration associated with a characteristic GW signal. 

\begin{figure}
\hspace*{-0.35cm}
\centering
\includegraphics[width=9cm,clip]{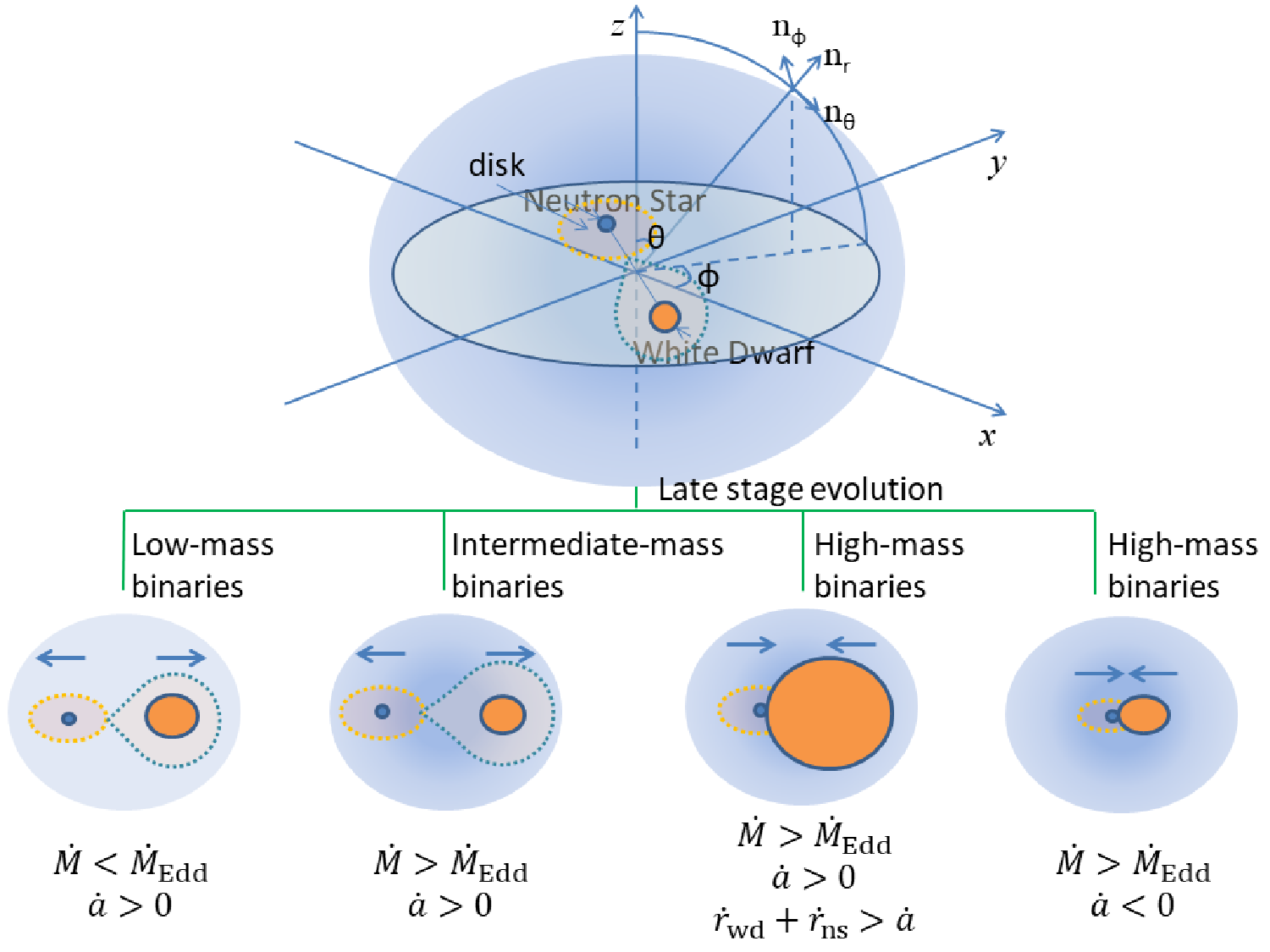}
\caption{
Schematic illustration of the polymorphic late-stage evolution of compact NSWD binaries with varying component masses. 
The diagram shows four distinct evolutionary outcomes depending on the mass transfer rate (\( \dot{M} \)), the Eddington accretion rate (\( \dot{M}_{\rm Edd} \)), the orbital separation (\( a \)), and its rate of change (\( \dot{a} \)). 
The spherical coordinate system \((r, \theta, \phi)\) is defined with basis vectors \((n_{\rm r}, n_{\theta}, n_{\phi})\), and the orientation of the binary orbit is described by the polar and azimuthal angles \((\theta, \phi)\) in the Cartesian frame \((x, y, z)\). 
The rates of change of the WD and NS radii are denoted by \( \dot{r}_{\rm wd} \) and \( \dot{r}_{\rm ns} \), respectively.
}
\label{fig1}
\end{figure}

\begin{figure}
\hspace*{-0.3cm}
\centering
\includegraphics[width=9cm,clip]{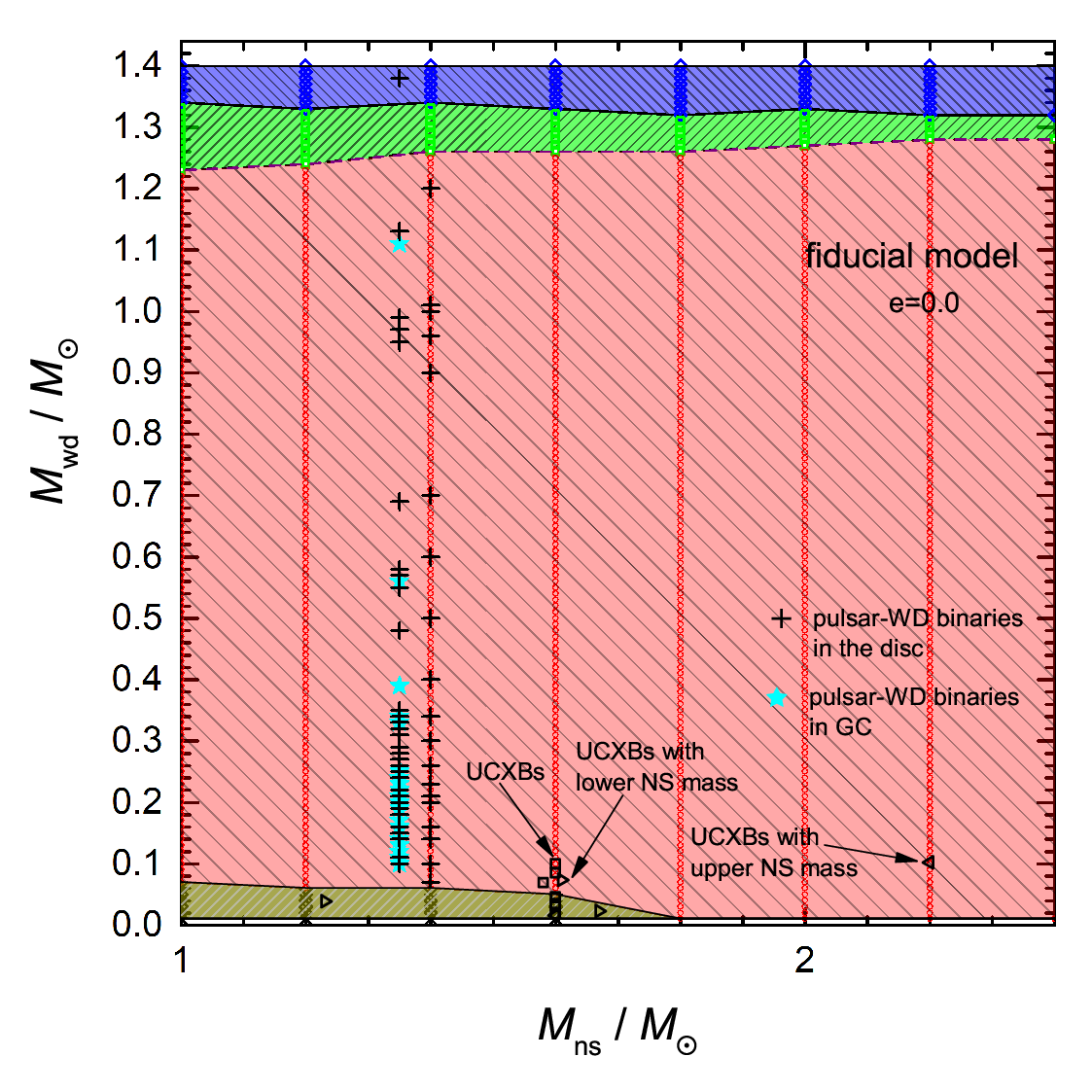}
\caption{
Distribution in \( M_{\rm ns} \)$-$\( M_{\rm wd} \) (neutron star mass$-$white dwarf mass) parameter space of polymorphic evolution channels 
in the late-stage evolution of circular NSWD binaries (\( e = 0.0 \)).  
Coloured regions represent the four distinct evolutionary outcomes: G1 (dark yellow), G2 (red), G3 (green), and G4 (blue). 
Computation grid points are indicated by crosses, red circles, green squares, and blue diamonds. 
Solid and dashed black lines delineate the boundaries between these groups.
Observational data are overlaid: pulsar-white dwarf binaries in the Galactic disk and globular clusters are marked with $`+`$ and $`\star`$ symbols, respectively, 
while ultracompact X-ray binaries are shown as open black squares and triangles. 
}
\label{fig2}
\end{figure}

\begin{figure*}
\centering
\includegraphics[width=0.95\textwidth,clip]{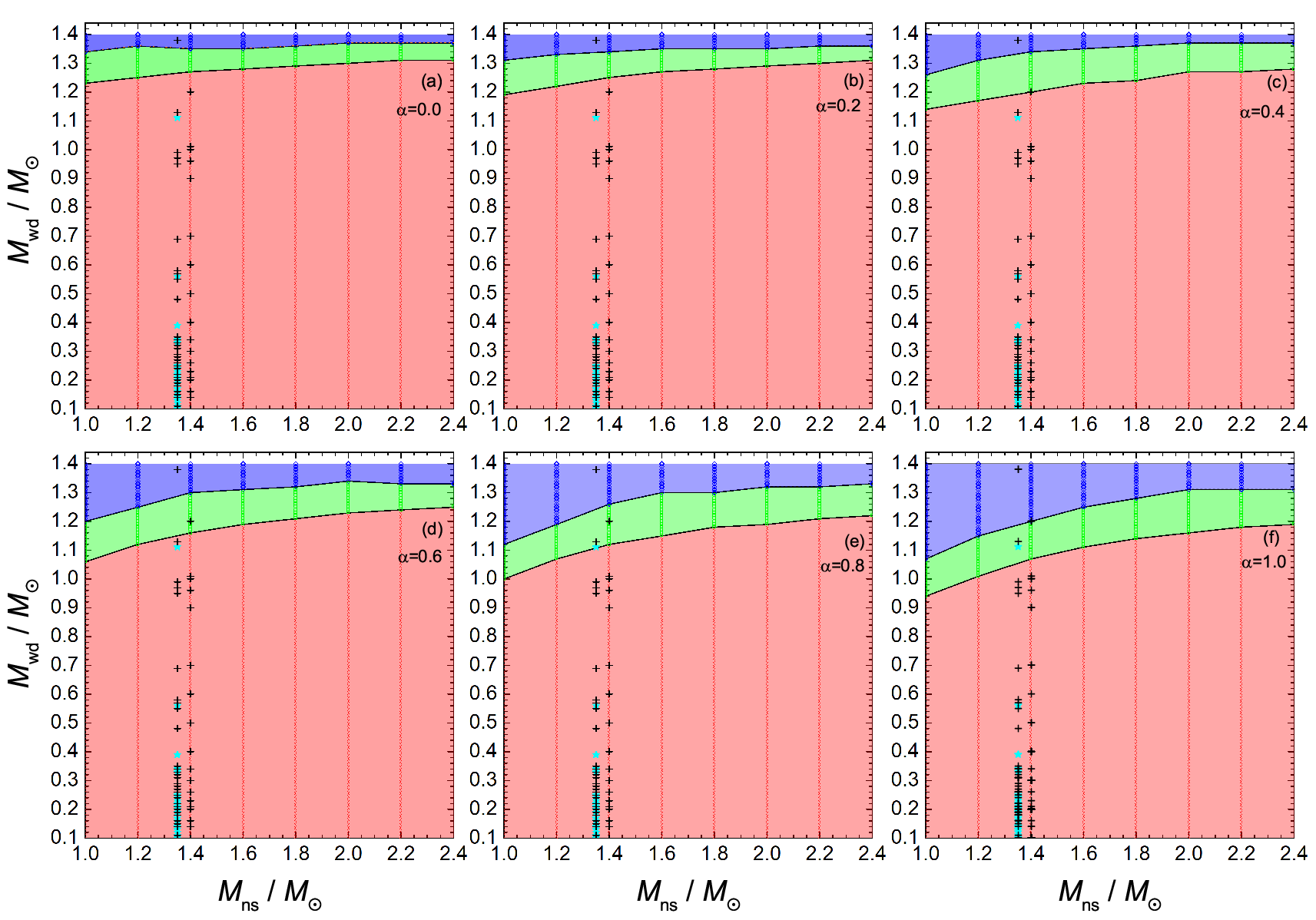}
\caption{
As Fig.~\ref{fig2}, but for different fixed values of the accretion fraction parameter \( \alpha \). 
Panels (a)-(f) correspond to \( \alpha = 0.0, 0.2, 0.4, 0.6, 0.8, \) and \( 1.0 \), respectively. 
All computations assume circular orbits (\( e = 0 \)). 
This figure illustrates how varying accretion efficiency influences the evolutionary outcomes of NSWD binaries.
}
\label{fig3}
\end{figure*}

\section{Method}
\label{sec_method}

\subsection{The onset of mass transfer}
Significant mass transfer occurs if the surface of the WD reaches its Roche lobe, 
when a neutron star and a white dwarf orbit each other. 
The WD will then expand as a result of losing mass. 
The binary system evolves from an inspiral phase driven by GW radiation to a mass transferring phase. 
The condition for the onset of mass transfer can be written as 
$r_{2}=r_{\rm 2L}$ where $r_{2}$ is the radius of white dwarf and $r_{\rm 2L}$ its Roche lobe radius. 
We estimate $r_{\rm 2L}$ according to the following approximation \citep{Eggleton83}
\begin{equation}
r_{\rm 2L}=a\frac{0.49q^{2/3}}{0.6q^{2/3}+\rm
ln(1+\it q^{\rm 1/3} )},
\label{eq_rl}
\end{equation}
where $q$ ($= m_{2}/m_{1}$) is the mass ratio, 
$m_{1}$ and $m_2$ the masses of the primary (NS) and secondary (WD) components.

\subsection{The WD and NS models}

In order to understand how mass transfer affects the orbital evolution of a NSWD binary, 
we execute a time-dependent numerical calculations to model the structure evolution of a WD and thus compute 
the evolution of mass transfer. 
In contrast to previous studies (e.g., \citet{Verbunt88,Marsh04,Paschalidis2009}), we assume that the 
self-gravity of a WD is balanced by the electron degenerate pressure $P_{\rm e}$, therefore the structure of a WD 
along its radius $r_{2}$ can be numerically obtained by solving the hydrostatic equation and Poisson’s equation:
\begin{equation}
\frac{{\rm d} P_{\rm e}}{{\rm d} r_{2}}=-\frac{{\rm d} \Phi}{{\rm d} r_{2}}\rho_{2},
\label{eq_hydrostatic}
\end{equation}
\begin{equation}
\nabla^{2} \Phi= 4\pi G \rho_{2},
\label{eq_poisson}
\end{equation}
where $\Phi$ is the gravitational potential, $\rho_{2}$ mass density of a WD and $G$ the gravitational constant. 
The degenerate pressure of an electron fermion gas $P_{\rm e}$ and the mass density of a WD can be written as \citep{Kippenhahn90}
\begin{equation}
P_{\rm e}=\frac{\pi m_{\rm e}^{4}c^{5}}{3h^{3}}\cdot \left[(2x^{3}-3x)\sqrt{x^{2}+1}+3 \ln (x+\sqrt{x^{2}+1})\right],
\label{eq_gpe}
\end{equation}
and
\begin{equation}
\rho_{2}=\mu_{\rm e}m_{\rm u} \frac{8\pi m_{\rm e}^{3}c^{3}}{3 h^{3}}x^{3}.
\label{eq_denm}
\end{equation}
where $m_{\rm e}$ is the rest mass of an electron, $m_{\rm u} \approx 1.66\times10^{-24}$\,g the atomic mass unit, $h$ the Planck constant, $x=p_{\rm m}/m_{\rm e}c$ with $p_{\rm m}$ being a given momentum. All phase cells below the momentum $p_{\rm m}$ are occupied by two electrons, and all phase cells above $p_{\rm m}$ are empty. For a pure electron gas, the mean weight $\mu_{\rm e}$ is
\begin{equation}
\mu_{\rm e} \approx \left(X+\frac{1}{2}Y+\frac{1}{2}Z\right)^{-1},
\label{eq_meanwe}
\end{equation}
where $X$, $Y$, and $Z$ represent the fraction of the mass density contributed by hydrogen, helium, 
and heavier elements in a star, respectively. We have $X+Y+Z=1$.

The mass fraction which flows outside of the Roche lobe per unit time, i.e., the mass transfer rate and mass loss from the binary system, can be roughly estimated as the product of volume ($\approx r_{2}^{3}-r_{\rm 2L}^{3}$) of overflow matter outside of the Roche lobe and mass density $\rho_{2}$ of the overflow matter, i.e.,
\begin{equation}
\dot{m}_{2}\approx \frac{\Delta m_{2}}{\Delta t}=-\frac{4\pi}{3}\frac{\rho_{2}(r_{2}^{3}-r_{\rm 2L}^{3})}{\Delta t},
\label{eq_mtrate}
\end{equation}
where $\Delta t$ is a small time interval (time step in the simulations).
The change of $r_{2}$ and $\rho_{2}$ with respect to time are obtained by
numerically solving Eqs.~\eqref{eq_hydrostatic}$-$\eqref{eq_meanwe}.

A simplified approach is employed in this paper to correct the deformation of WD 
caused by tidal interaction between the WD and NS. 
We refer to the shape deformation of WD as the tidal bulge.
The height of the bulge referred to the mean radius is about \citep{goldreich66}:
\begin{equation}
\begin{split}
&\Delta r_{2}=\epsilon\frac{3}{4}\frac{m_{1}r_{2}^{4}}{m_{2}a^{3}},\\
&\epsilon=\frac{5}{2+19\iota_{2}/g_{2s}\rho_{\rm 2s} r_{2}},
\label{eq_bulgeheighttide}
\end{split}
\end{equation}
where $\epsilon$ is the correction factor for the rigidity of the white dwarf and
for a second degree disturbance of the tidal potential due to the deformation.
$\iota_{2}$, $g_{2s}$ and $\rho_{\rm 2s}$ represent the rigidity, surface
gravity, and tidal bulge density of a white dwarf respectively \citep{goldreich66}.
For WDs, we assume $19\iota_{2}\lesssim g_{2s}\rho_{\rm 2s} r_{2}$ which gives the maximum value
$\epsilon_{\rm max}=5/2$.
In order to simplify our model in this paper, we take $\epsilon=\epsilon_{\rm max}$ as a constant
to reduce the number of free parameters.
In due course, the rigidity parameter $\iota_{2}$ and its dependence on the interior physics of the white dwarf needs to be investigated in more detail.
The influence of the tidal deformation on the WD radius in our model can be found in \citet{Yu21}.

Recent multimessenger data analysis places constraints on the maximal NS mass 
to be in the range of $\sim$2-2.6$M_{\odot}$ \citep{Rezzolla2018,Margalit2017,Most2020,Godzieba2021,Li2021,Shibata2019,Xue2025}. 
\citet{Ozel2016b} have employed Bayesian statistical frameworks to obtain NS radii 
from spectroscopic measurements as well as to infer the Equation of State (EoS) from the radius measurements. 
Combining these with the results of experiments in the vicinity of nuclear saturation density and the
observations of $\sim2M_{\odot}$ NSs, they have imposed constraints on the properties of the EoS 
between $\sim$(2-8)$\rho_{\rm nm}$ ($\rho_{\rm nm}$: nuclear matter saturation density). 
They found that around neutron star mass $M_{\rm ns}=1.5M_{\odot}$ the preferred EoS predicts 
radii between 10.1 and 11.1 km. 

\citet{Fan2024} explored the mass-tidal deformability data of GW170817, the mass-radius data of pulsars, 
as well as the theoretical calculations from the chiral effective theory and perturbative quantum chromodynamics. 
They suggested 2.25$^{+0.08}_{-0.07}M_{\odot}$ (68.3\% credibility) to be the maximal gravitational mass of 
nonrotating neutron stars and 11.90$^{+0.63}_{-0.60}$km to be the radius of the most massive nonrotating neutron star. 
The stiff EoS models with $M_{\rm TOV} \geq 2.4M_{\odot}$ is disfavored at a confidence level above 95\% 
in their calculations. 

To simplify our model, we in this paper do not consider the EoS of neutron stars, but assume a NS radius 
to be 10km when the NS mass is in the range of $1.0-2.4M_{\odot}$. The assumption of 10km of NS radius in 
our models is 
a main difference compared to the point mass assumption of NS in previous studies (e.g. \citet{Paschalidis2009,Zenati19,Moran-Fraile2024}). 
This assumption allows us to calculate the mass loss from a binary system powered by 
the gravitational potential energy released from the accretion process of NS, 
which will be presented in the following section.  
 
\subsection{The fraction of accreted matter}

During mass transfer, the donor star \( m_2 \) loses mass at a rate \( -\dot{m}_2 \). 
A fraction \( \alpha \dot{m}_2 \) is accreted by the neutron star, while the remaining 
\( (1 - \alpha) \dot{m}_2 \) is lost from the system. 
\( \alpha \) is the fraction of the white dwarf's lost mass that is accreted by the neutron star.
Thus, the total mass change rate is \( \dot{M} = (1 - \alpha) \dot{m}_2 \). 
In the limiting cases, \( \alpha = 1 \) corresponds to fully conservative mass transfer, 
while \( \alpha = 0 \) represents fully non-conservative evolution.

In our fiducial model, \( \alpha \) is dynamically determined based on the accretion luminosity: 
\begin{enumerate}
    \item If the luminosity is below the Eddington limit \( L_{\rm Edd} \), all transferred mass is accreted.
    \item If the luminosity exceeds \( L_{\rm Edd} \), and the photon diffusion timescale is longer than the inward advection timescale, 
    excess energy can be absorbed in the overflow mass and will drive some of the mass flow away from the boundary of Roche lobe 
    into a common envelope. 
    The value of \( \alpha \) is then derived from energy conservation \citep{Han99,Yu21}.
\end{enumerate}

We also calculate the NSWD evolution in the cases of $\alpha=0.0,0.2,0.4,0.6,0.8,1.0$ for comparison. 

\subsection{Orbital Evolution}

We compute the orbital evolution of NSWD binaries by evaluating the time derivative of the orbital separation, \( \dot{a} \), through the evolution of the orbital angular momentum, \( \dot{J}_{\rm orb} \). 
This evolution is governed by three primary mechanisms: gravitational wave radiation (\( \dot{J}_{\rm GR} \)), mass transfer (\( \dot{J}_{\rm mt} \)), and spin-orbit coupling between the accretor and the binary orbit (\( \dot{J}_{\rm so} \)) \citep{Yu21}.

When RLOF occurs, we solve the hydrostatic and Poisson equations to determine the internal structure of the white dwarf. 
The mass transfer rate is then estimated based on the volume and density of the overflowing material.

To model GW emission, we calculate the GW strain in the direction \( (\theta, \phi) \) in the spherical coordinate system \( (n_{\rm r}, n_{\theta}, n_{\phi}) \) of the source frame. 
The GW amplitudes for the plus and cross polarisations are given by:

\begin{equation}
\begin{split}
h_{+} = \frac{G}{3R_{\rm b}c^{4}} \Big[ & \ddot{A}_{\rm xx}(\sin^{2}\phi - \cos^{2}\phi \cos^{2}\theta) \\
+ & \ddot{A}_{\rm yy}(\cos^{2}\phi - \sin^{2}\phi \cos^{2}\theta) \\
- & \ddot{A}_{\rm zz} \sin^{2}\theta \\
- & \ddot{A}_{\rm xy} \sin(2\phi)(1 + \cos^{2}\theta) \\
+ & \ddot{A}_{\rm xz} \cos\phi \sin(2\theta) \\
+ & \ddot{A}_{\rm yz} \sin\phi \sin(2\theta) \Big],
\end{split}
\label{eq_hplus}
\end{equation}

\begin{equation}
\begin{split}
h_{\times} = \frac{G}{3R_{\rm b}c^{4}} \Big[ & (\ddot{A}_{\rm xx} - \ddot{A}_{\rm yy}) \sin(2\phi) \cos\theta \\
- & 2\ddot{A}_{\rm xy} \cos(2\phi) \cos\theta \\
- & 2\ddot{A}_{\rm xz} \sin\phi \sin\theta \\
+ & 2\ddot{A}_{\rm yz} \cos\phi \sin\theta \Big],
\end{split}
\label{eq_hcross}
\end{equation}

where \( \ddot{A}_{\alpha\beta} \) denotes the second time derivative of the mass quadrupole tensor in Cartesian coordinates \( (x, y, z) \), \( G \) is the gravitational constant, \( c \) is the speed of light, and \( R_{\rm b} \) is the distance to the observer. 
These expressions are consistent with standard formulations in the literature \citep{Yu15, Maggiore08}.

We perform simulations over a large parameter grid of size \( 8 \times 136 \times 5 \times 6 \), 
spanning neutron star mass, white dwarf mass, orbital eccentricity, and mass accretion efficiency \( \alpha \). 
The initial values of the four parameters are listed in Table~\ref{tab1}. 
The initial orbital separation is adopted to be slightly greater than that when a WD 
is about to fill its Roche lobe so as to calculate inspiral and mass transferring phase 
in a time-efficient manner.

\begin{table}
\caption{The ranges and increments of four main initial parameters used in the computational grid.}
\label{tab1}
\centering
\begin{tabular}{lcc}
\hline\hline
Parameter & Range & Increment \\
\hline
$M_{\rm ns}$ (Neutron star mass) & 1.0--2.4 $\Msolar$ & 0.2 $\Msolar$ \\
$M_{\rm wd}$ (White dwarf mass) & 0.05--1.4 $\Msolar$ & 0.01 $\Msolar$ \\
$e$ (Eccentricity) & 0.0--0.8 & 0.2 \\
$\alpha$ (Accretion fraction) & 0.0--1.0 & 0.2 \\
\hline
\end{tabular}
\end{table} 
 
\subsection{GW emission from accretion disks}
\label{sec_disk}

To estimate the GW emission from an accretion disk, we divide the disk into \( N \) azimuthal segments, each treated as a point mass. 
Each segment and the neutron star form an ``element binary'', for which we compute the evolution of the quadrupole moment. 
The total GW amplitude from the disk is obtained by coherently summing the contributions from all segments:

\begin{equation}
\begin{split}
h_{\rm disk} = \sum_{n=1}^{N} \tilde{h}_{n}(\theta, \phi_{n}), \quad \phi_{n} = \frac{2\pi n}{N},
\end{split}
\end{equation}
where \( \tilde{h}_{n}(\theta, \phi_{n}) \) is the GW amplitude from the \( n \)-th element binary.

We assume that the disk extends from the innermost stable circular orbit (ISCO) to the L1 point of the white dwarf. 
Beyond the ISCO, stable orbits are no longer possible in Schwarzschild geometry. The ISCO radius is 
approximated as \( r_{\rm ISCO} = 6GM/c^{2} \approx 8.9(M/\Msolar)~\mathrm{km} \). 

\subsection{Dynamical Power Spectral Density}
In order to obtain the dynamical power spectral density (DPSD) \citep{Press92}, we execute the following procedure: \\
1. Calculate an NSWD binary evolution in time domain using time-step $\Delta t_{0}$, and take M$\times$(N-point sample) of the GW amplitude $h(t)$
at equal time intervals $\Delta t_{1}$;   \\
2. Divide the M$\times$(N-point sample) into M data sets, so that there are N points in each data set; \\
3. Apply the Discrete Fourier Transform (DFT) to the first data set, so the frequencies 
at the $k$th data point in this data set are $f_{\rm k}=\frac{k}{N\cdot\Delta t_{1}}$, $k=$0, 1, ..., $\frac{\rm N}{2}$; \\
4. Apply the DFT to the second data set, ..., and to the Mth data set. \\ 

Note that we apply square windowing in our calculations, there is no noise being injected, and the spectra are not whitened.  

\section{Results}
\label{sec_results}

The GW emission from a NSWD binary can be approximated by treating the components as point masses, provided the Roche lobe radius of each star is significantly larger than its physical radius. 
This approximation holds during the early in-spiral phase, before the onset of Roche lobe overflow (RLOF). 
However, once the white dwarf fills its Roche lobe and mass transfer begins, the system can no longer be treated as a pair of point masses. 
In this regime, the GW signal becomes closely linked to the system's dynamic chirp mass \citep{Tauris18}.

We compute the GW waveforms of NSWD binaries by modelling their evolution through RLOF. 
Our results show that the waveforms are highly sensitive to the initial component masses and the surrounding environment. 
By varying environmental parameters while keeping the initial masses fixed, we demonstrate that NSWD binaries with the same initial configuration \textit{may} follow distinct evolutionary paths-an effect we refer to as \textit{branched} or \textit{polymorphic evolution}. 
This leads to diverse GW signatures even among systems with identical initial masses.

\subsection{The evolution of  mass-transferring neutron star-white dwarf binaries}

Mass transfer plays a critical role in shaping the GW signal during the RLOF stage, as the emission is governed by the second time derivative of the system's quadrupole moment. 
We simulate the orbital evolution of a Roche-lobe-filling NSWD binary and compute its quadrupole moment accordingly. 
The influence of an accretion disk on the GW signal is also incorporated.
The orbital separation \( a \) of a Roche-lobe-filling NSWD binary evolves using 
the results in \citet{Yu21}: 

\begin{equation}
\frac{\dot{a}}{2a} = -\frac{32}{5}\frac{G^{3}}{c^{5}}\frac{m_{1}m_{2}M}{a^{4}} + f(q) \frac{-\dot{m}_{2}}{m_{2}} + \dot{J}_{\rm so},
\end{equation}

\begin{equation}
f(q) = 1 - \alpha q - \left(1 - \alpha\right) \frac{q}{1 + q} \left(\beta + \frac{1}{2}\right) - \alpha \sqrt{(1 + q) r_{\rm h}},
\label{eq_fq}
\end{equation}

where:
\begin{itemize}
    \item \( M = m_1 + m_2 \) is the total mass,
    \item \( \dot{J}_{\rm so} \) represents the angular momentum exchange between the accretor's spin and the binary orbit.
    For the NSWD binary with sufficiently long synchronization time-scale, the term \( \dot{J}_{\rm so} \) is smaller than 
    the first two items \citep{Yu21}. We neglect this term in this paper.    
    \item \( \beta = 1 \) is the specific angular momentum loss parameter,
    \item \( r_{\rm h} \) is the dimensionless radius associated with the accretion disk \citep{Lubow75, Verbunt88}. 
\end{itemize}

The function \( f(q) \) encapsulates the effects of mass transfer and angular momentum loss.
Since \( -\dot{m}_2 / m_2 > 0 \), the sign of \( f(q) \) determines the system's fate: 
if \( f(q) < 0 \), the orbital separation shrinks continuously, leading to coalescence; 
if \( f(q) > 0 \), the system may avoid merger, depending on the mass loss rate. 
In Eq.\,\ref{eq_fq}, the terms represent, respectively, the effects of mass changes 
in the white dwarf and neutron star, systemic mass loss, and the influence of the accretion disk.

\subsection{Conservative evolution}
\label{sec_ce}

In the conservative mass transfer scenario, all material lost by the white dwarf (WD) is accreted by the neutron star (NS), i.e., \( \alpha = 1 \). 
Under this condition, the function \( f(q) \) simplifies to:
\begin{equation}
f(q) = 1 - q - \sqrt{(1 + q) r_{\rm h}},
\end{equation}
where \( q = m_2 / m_1 \) is the mass ratio and \( r_{\rm h} \) characterises the size of the accretion disk. In this regime, \( f(q) \) is primarily governed by the mass ratio and the disk size.

For systems with relatively massive WDs (i.e., \( q \approx 1 \)), achieving \( f(q) < 0 \) requires a compact accretion disk. 
As the system evolves and mass is transferred from the WD to the NS, the mass ratio \( q \) decreases, eventually leading to \( f(q) > 0 \) beyond a critical point. 
This transition marks a shift in the orbital evolution behaviour.

\subsection{Completely non-conservative evolution}
\label{sec_cne}

In the fully non-conservative case, where no mass is accreted by the NS (\( \alpha = 0 \)), the function \( f(q) \) becomes:
\begin{equation}
f(q) = 1 - \frac{q}{1 + q} \left( \beta + \frac{1}{2} \right),
\end{equation}
where \( \beta \) represents the specific angular momentum carried away by the ejected mass. 
Here, the evolution of \( f(q) \) is determined by the mass ratio \( q \) and the angular momentum loss.

For the binary to coalesce (\( f(q) < 0 \)), the condition \( q > \frac{2}{2\beta - 1} \) must be satisfied. 
In the commonly adopted case of \( \beta = 1 \), this implies \( q > 2 \), which contradicts the typical range \( 0 < q \leq 1.4 \). 
Therefore, in this regime, whether the binary merges depends more critically on the magnitude of the WD's mass loss rate than on angular momentum loss alone.

\subsection{Mergers}
\label{sec_mergers}

Based on the mass transfer state and the likelihood of coalescence, NSWD binary evolution can be categorised into four distinct groups:

\begin{itemize}
    \item \textbf{G1: Non-coalescing systems with sub-Eddington Accretion (Low-Mass Mergers)} \\
    Systems with mass transfer rates below the Eddington limit evolve on long timescales (\textit{e.g.} 
    a binary with white dwarf mass \( M_{\rm wd} \lesssim 0.03~\Msolar \) and \( M_{\rm ns} = 1.4~\Msolar \) ). 
    These ultra-low-mass mergers are not considered further in this study.

    \item \textbf{G2: Non-coalescing systems with super-Eddington Accretion and envelope expansion (Intermediate-Mass Mergers)} \\
    Systems in which mass transfer exceeds the Eddington limit will form a common envelope. 
    Initially, the orbital separation decreases due to GW radiation, but later increases as mass transfer dominates. 
    If the orbital expansion outpaces the WD's expansion, the binary avoids coalescence. 

    \item \textbf{G3: Expansion-Induced Coalescence (High-Mass Mergers)} \\
    If the WD's radius increases faster than the orbital separation, the binary will evolve into contact and eventual merger. 

    \item \textbf{G4: Direct Coalescence (High-Mass Mergers)} \\
    In high-mass (and high $\dot{M}$) systems, the WD radius increases so quickly that the WD plunges directly into the NS without significant orbital expansion. 

    The G3 and G4 groups are collectively called \textit{high-mass mergers}. 
    Typically, in these cases, we have \( M_{\rm wd} \gtrsim 1.22~\Msolar \) in the fiducial model, 
    and \( M_{\rm wd} \gtrsim 0.94~\Msolar \) in the case of conservative evolution ($\alpha=1$).
   
    For the low-mass and intermediate-mass mergers (non-coalescing systems), 
    the mass transfer process dominates the orbital evolution, 
    while for the high-mass mergers (coalescing systems) the dominant process is the GW radiation. 
\end{itemize}

This classification is illustrated in Fig.~\ref{fig1}, along with the coordinate systems used in the source frame. 
The mass boundaries separating coalescing and non-coalescing systems under various evolutionary conditions are shown in Figs.~\ref{fig2}--\ref{fig3}. 
Figure~\ref{fig2} presents results from our fiducial model, where \( \alpha \) is dynamically determined.

To assess the impact of different accretion efficiencies, we fix \( \alpha \) between 0.0 and 1.0 in panels (a)--(f) of Fig.~\ref{fig3}. 
Our results show that eccentricity can shift the mass boundaries of merging systems by up to \(\sim 10\%\).
However, high eccentricity may also alter the contact state and deform the WD on timescales comparable to half the orbital period.

For \( \alpha \lesssim 0.4 \), the WD mass boundary for mergers lies between \( 1.1 \) and \( 1.3~\Msolar \), largely independent of the NS mass. When \( \alpha \gtrsim 0.6 \), the WD mass threshold increases with NS mass. 
For example, with \( \alpha = 1.0 \) and \( M_{\rm ns} = 1.0~\Msolar \), the minimum WD mass for merger is \( 1.07~\Msolar \) in the direct coalescence case and \( 0.94~\Msolar \) in the radius expansion case. 
For \( M_{\rm ns} = 2.4~\Msolar \), these thresholds rise to \( 1.31~\Msolar \) and \( 1.19~\Msolar \), 
respectively. 

Comparing Figs.~\ref{fig2} and \ref{fig3}, we find that in our fiducial model, \( \alpha \) typically evolves to \( \lesssim 0.2 \). This implies that only a small fraction of the WD's lost mass is accreted by the NS, forming a low-mass disk, while most of the mass is expelled into a common envelope or beyond. For comparison, observed NSWD binary masses are also plotted in Figs.~\ref{fig2}--\ref{fig3}. 

We use a polynomial equation $m_{2}=\eta_{0}+\eta_{1}m_{1}+\eta_{2}m_{1}^{2}+\eta_{3}m_{1}^{3}$ 
to fit the relation of WD mass $m_{2}$ and NS mass $m_{1}$ for the boundary 
of coalescing (G3+G4) and non-coalescing system (G1+G2) in the cases of $\alpha=0.0-1.0$. 
The coefficients $\eta_{0}-\eta_{3}$ and the adjusted R-squared are listed in Table\,\ref{tab2}. 
Note that in these fitting equations, we assume the eccentricity $e=0$, and the eccentricity may 
affect the mass boundaries of coalescing and non-coalescing systems by $\lesssim10\%$.

\begin{table*}
\caption{
We use a polynomial equation $m_{2}=\eta_{0}+\eta_{1}m_{1}+\eta_{2}m_{1}^{2}+\eta_{3}m_{1}^{3}$ 
to fit the relation of WD mass $m_{2}$ and NS mass $m_{1}$ for the boundary 
of coalescing (G3+G4) and non-coalescing system (G1+G2) in the cases of $\alpha=0.0-1.0$. 
The coefficients $\eta_{0}-\eta_{3}$ and the adjusted R-squared are listed in this table. 
} 
\label{tab2}
\begin{center}
\fontsize{9}{9}\selectfont{
\begin{tabular}{lccccc}
\hline\hline
                       &    $\eta_{0}$    &   $\eta_{1}$     &  $\eta_{2}$    &   $\eta_{3}$    &  adjusted R-squared       \\
 \hline 
 fiducial model        & 1.0532            & 0.2947   & -0.143  & 0.0252 & 0.90944 \\
 $\alpha=0.0$          & 1.0680            & 0.2208   & -0.0649 & 0.0063 & 0.99255 \\
 $\alpha=0.2$          & 0.8097            & 0.5992   & -0.2613 & 0.0410 & 0.99675 \\
 $\alpha=0.4$          & 0.955             & 0.188    & 0.007   & -0.012 & 0.9873  \\
 $\alpha=0.6$          & 0.4652            & 0.9239   & -0.385  & 0.0568 & 0.99898 \\
 $\alpha=0.8$          & 0.236             & 1.213    & -0.531  & 0.082  & 0.99799 \\
 $\alpha=1.0$          & 0.22814           & 1.06921  & -0.4147 & 0.05682 & 0.99943 \\
\hline
\end{tabular}
}
\end{center}
\end{table*}

\subsection{Gravitational wave morphology}
\label{sec_pgw}

We illustrate the diverse evolution morphology of GW signals produced by four NSWD binaries, each evolved from known pulsar-white dwarf systems identified through radio observations. 
These examples are analysed under both our fiducial model and the conservative mass transfer scenario. 
The component masses, eccentricities, and orbital periods of the progenitor systems are derived from observational data, 
which are listed in Table\,\ref{tab3}.

\begin{figure*}
\hspace*{-0.0cm}
\centering
\includegraphics[width=0.95\textwidth,clip,angle=0]{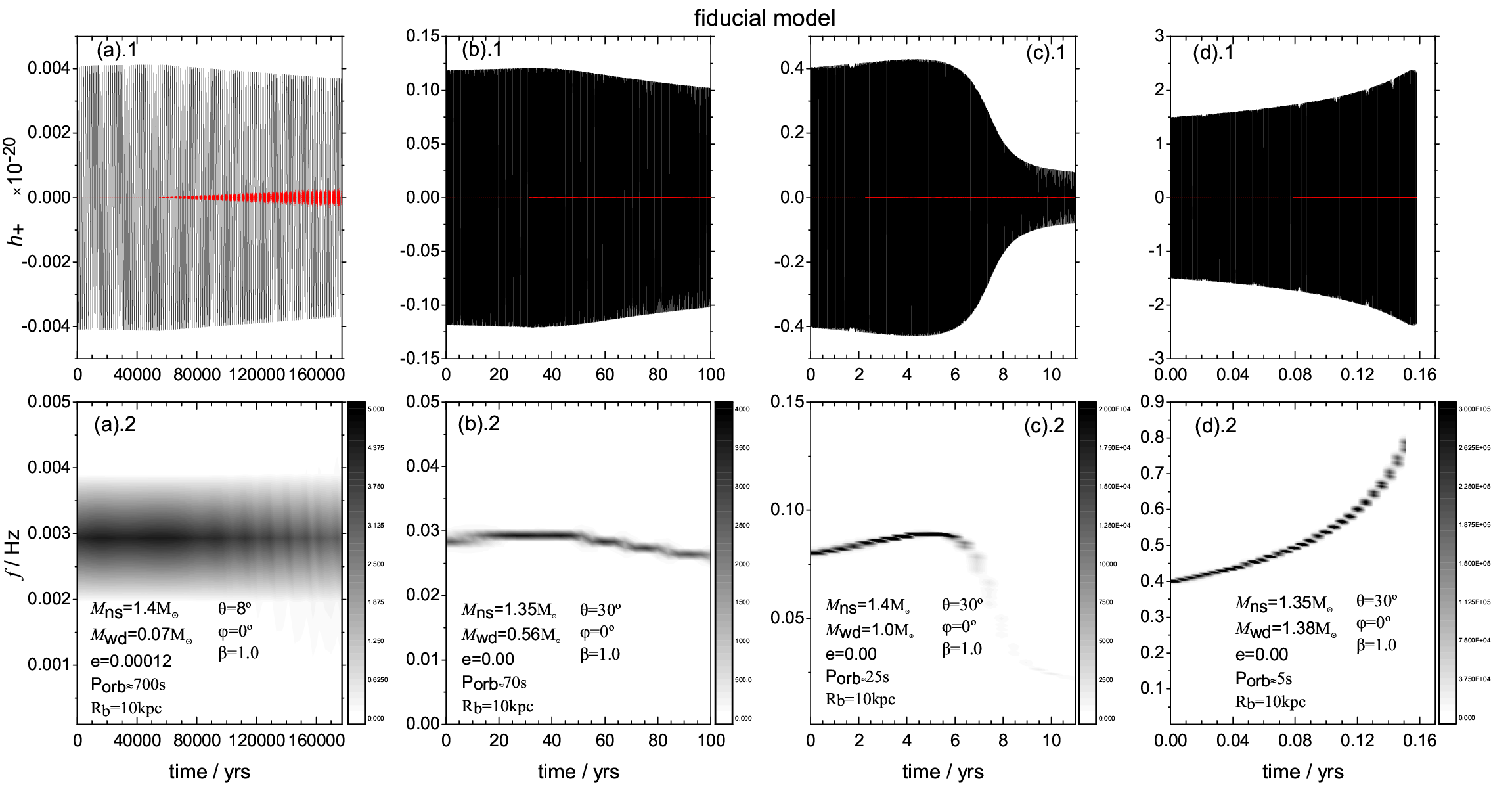}
\caption{
The evolution of gravitational wave (GW) signals generated by four NSWD binaries, each evolved from known pulsar--white dwarf systems, modelled under our fiducial assumptions. We assume that the pulsar component in each binary is a rapidly rotating neutron star. 
Top and bottom panels (labeled $\ast.1$ and $\ast.2$) show the evolution of the GW plus polarisation \( h_+ \) and its corresponding dynamic power spectral density (DPSD), respectively. The initial parameters for each binary are indicated in the lower panels. 
In each top panel, the red dotted line represents the GW signal contribution from the accretion disk surrounding the neutron star.}
\label{fig4}
\end{figure*}

\begin{figure*}
\hspace*{-0.0cm}
\centering
\includegraphics[width=0.95\textwidth,clip,angle=0]{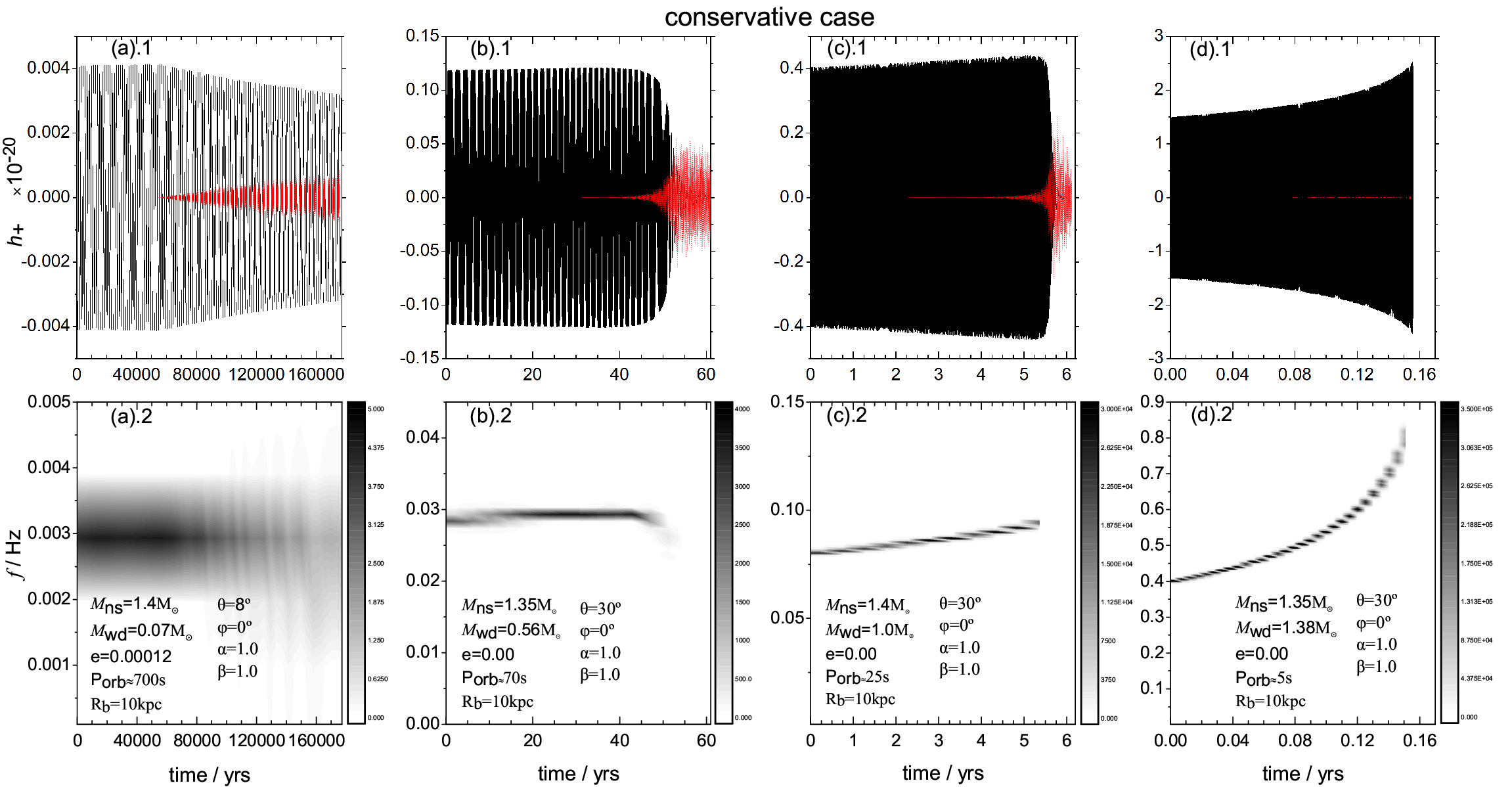}
\caption{
As Fig.\,\ref{fig4}, except that all the mass loss from the WD component is assumed to be accreted by the NS component.
}
\label{fig5}
\end{figure*}

We assume that each binary will eventually enter the RLOF phase due to orbital decay driven by GW radiation, neglecting other potential influences on orbital evolution. 
For each case, we simulate the system at the onset of RLOF. Figures~\ref{fig4} and \ref{fig5} present the resulting GW strain (top panels) and the corresponding DPSD 
(bottom panels). 
In the DPSD calculations, we take window length 1024 in Fig.\,\ref{fig4}. 
For panels (a)-(d) in Fig.\,\ref{fig4}, we adopt time-step $\Delta t_{0}=4,~0.004,~5\times10^{-4},~5\times10^{-6}$ yrs, 
and time interval $\Delta t_{1}=1,~1,~1,~0.2$ s, respectively. 

Panels (a)-(d) in Figs.~\ref{fig4}-\ref{fig5} reveal a clear trend: low-mass NSWD binaries exhibit nearly flat-frequency DPSD profiles over long timescales (cases G1-G2), while high-mass systems show rapidly increasing frequencies over shorter timescales (G4). 
Although the accretion efficiency parameter \( \alpha \) has little effect on the peak amplitude of \( h_{+} \), it significantly alters the waveform's temporal evolution.

Observations from LISA-type detectors of harmonic GW signals in high-eccentricity binaries may provide insights into harmonic generation and white dwarf deformation.

The binary in Panel (a) is likely to evolve into an ultracompact X-ray binary, potentially undergoing mildly super-Eddington accretion (case G2).  
Intermediate-mass systems (Panels b and c) emit GW signals with amplitudes around \( \sim 10^{-21} \), accompanied by multi-wavelength electromagnetic radiation, as mass lost to the common envelope is heated by high-energy photons and gravitational potential energy (case G3).
The system in Panel (d) is a coalescing binary, producing GW signals up to \( \sim 2.5 \times 10^{-20} \), with the GW frequency increasing from approximately 0.4 Hz to 0.82 Hz over a two-month timescale (case G4).

In the conservative case, the accretion disk may also contribute to the GW signal. As shown by the red dotted lines in Figs.~\ref{fig4}-\ref{fig5}, disk-generated GW signals can reach amplitudes of \( \sim 10^{-22} \) to \( 10^{-21} \) as the accreted mass increases. However, in coalescing systems, even under conservative conditions, disk-related GW signals are typically absent due to the violent nature of the merger, which inhibits disk formation.

\begin{table*}
\caption{
Observed pulsar-white dwarf binaries expected to merge within \(10^{10}\) years, including three representative systems (out of eight total; see \protect \citep{Yu21}) and the high-mass system 
PSR J2222-0137, shown here for illustration. 
The table includes the following parameters: \(P\) - orbital period; \(e\) - eccentricity; \(t_{\rm m}\) - estimated merger timescale; \(m_1\) - assumed pulsar mass; \(m_2\) - estimated median companion mass assuming an inclination angle of \(60^\circ\), except for PSR J1953+1844 (M71E) which has an inclination of \(\sim 8^\circ\) \protect \citep{Pan23}; \(S\) - pulsar spin frequency; and \(d\) - distance to the system. 
All binary properties are sourced from the Australia Telescope National Facility (ATNF) Pulsar Catalogue (\url{http://www.atnf.csiro.au/research/pulsar/psrcat}) \protect \citep{Manchester05} and references therein.
}
\label{tab3}
\begin{center}
\fontsize{7}{7}\selectfont{
\begin{tabular}{lccccccc}
\hline
         Name (PSR)    &      $P$         & $e$               & $t_{\rm m}$       & $m_{1}$         & $m_{2}$        & S  &  $d$   \\
                       &                  &                   & ($\times10^{9}$ yr)     & ($\Msolar$)   & ($\Msolar$)  & (Hz)  & (kpc)  \\
 \hline

         J1953+1844 & 0.037$^{\rm d}\xrightarrow[]{t_{\rm m}}690^{\rm s}$ & 6.0E-4$\xrightarrow[]{t_{\rm m}}$1.2E-4 & $\sim$0.081 & 1.40 & 0.07 & 225.0& 4.81 \\
         J1141-6545       & 0.20$^{\rm d}\xrightarrow[]{t_{\rm m}}23.4^{\rm s}$ & 1.7E-1$\xrightarrow[]{t_{\rm m}}$0.00 & $\sim$0.53 & 1.40  & 1.00  & 2.539 & 3.20 \\
         J1748-2446N      & 0.39$^{\rm d}\xrightarrow[]{t_{\rm m}}67.7^{\rm s}$ & 4.5E-5$\xrightarrow[]{t_{\rm m}}$0.00	& $\sim$6.3  & 1.35 & 0.56  & 115.4  & 5.5  \\
         J2222-0137       & 2.45$^{\rm d}\xrightarrow[]{t_{\rm m}}4.1^{\rm s}$  & 3.8E-4$\xrightarrow[]{t_{\rm m}}$0.00 & $\sim$384  & 1.35  & 1.38  & 30.47 & 0.27 \\

\hline
\end{tabular}
}
\end{center}
\end{table*}

\subsection{Characteristic strain and signal-to-noise ratio}
\label{sec_snr}

We compute the sky-averaged squared signal-to-noise ratio (SNR), \( \overline{\rho^2} \), by averaging over source sky location, inclination, and polarisation angles \citep{Larson00,Moore15,LISA2018,Robson19}. 
This is approximated using the characteristic strain \( h_{\rm c} \) and the detector noise amplitude \( h_{\rm n} \), defined as:
\begin{equation}
h_{\rm c}(f) = \mathcal{A}(f) \sqrt{\frac{16}{5} f \cdot \Delta f}, \quad h_{\rm n}(f) = \sqrt{f \cdot S_{\rm n}(f)},
\end{equation}
where \( f \) is the GW frequency, \( \Delta f \) is the frequency evolution during the observation period, \( \mathcal{A}(f) \) is the Fourier amplitude of the GW signal computed from the time-domain waveform \( h_{+}(t) \), and \( S_{\rm n}(f) \) is the detector's effective noise power spectral density. 

For LISA, the power spectral density of total detector noise
$P_{\rm n}=\frac{1}{L^{2}}\left[P_{\rm o}+2(1+\cos^{2}(f/f_{\ast}))\frac{P_{\rm a}}{(2\pi f)^{4}}\right]$,
where $f_{\ast}=c/(2\pi L)$, $L=2.5\times10^{9}$ m is the armlength of the detector,
$P_{\rm o}=2.25\times10^{-22} ~\rm m^{2}\left(1+(\frac{2~mHz}{\it f})^{4}\right) ~~ \rm Hz^{-1}$
is the single-link optical metrology noise, and
$P_{\rm a}=9.0\times10^{-30} ~\rm (m~s^{-2})^{2}\left(1+(\frac{0.4~mHz}{\it f})^{2}\right)\left(1+(\frac{\it f}{8~\rm mHz})^{4}\right) ~~Hz^{-1}$
is the single test mass acceleration noise \citep{LISA2018, Robson19}.
$R(f)$ is the transfer function numerically calculated from \citet{Larson00}.
The effective noise power spectral density can be defined as $S_{\rm n}(f)=P_{\rm n}(f)/R(f)$.
For Taiji and Tianqin, we use the sensitivity curve data in \citet{Ruan20} and \citet{Wang19} respectively.
 
We also consider the deci-Hz detector B-DECIGO, with an arm length of approximately 100 km and a noise level of \( \sim 2.0 \times 10^{-23}~\mathrm{Hz}^{-1/2} \) near 1 Hz \citep{Isoyama18}.
The formulation allows us to estimate the SNR directly from a characteristic strain-frequency diagram. 

\begin{figure*}
\hspace*{-0.0cm}
\centering
\includegraphics[width=1.05\textwidth,clip,angle=0]{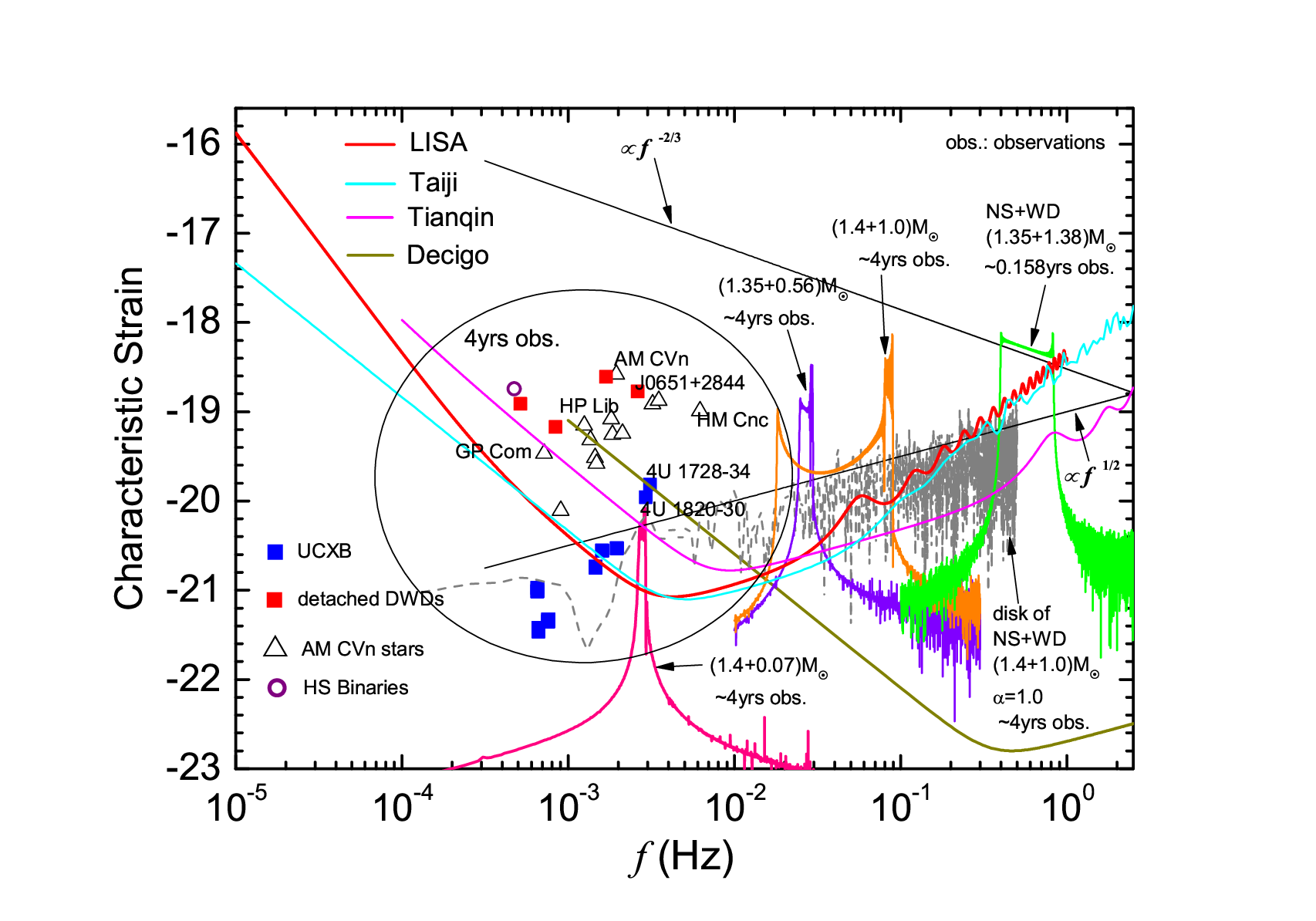}
\caption{We present the evolutionary tracks of four representative NSWD mergers, from the onset of Roche lobe overflow to the late stages of coalescence, in the characteristic gravitational wave (GW) strain–frequency diagram. 
The pink, purple, orange, and green curves correspond to systems with component masses of \(1.40~\Msolar + 0.07~\Msolar\), \(1.35~\Msolar + 0.56~\Msolar\), \(1.40~\Msolar + 1.00~\Msolar\), and \(1.35~\Msolar + 1.38~\Msolar\), respectively. Observation durations are indicated in the figure. 
Also shown are the characteristic strains of several known electromagnetic (EM) sources assuming a 4-year observation period: ultracompact X-ray binaries (blue squares), detached double white dwarfs (red squares), AM CVn systems (open triangles), and hot subdwarf (HS) binaries (purple open circles). 
The black lines represent reference spectra: a canonical in-spiral GW spectrum and a flat spectrum. Sensitivity curves for space-based GW detectors are plotted for comparison: LISA (red), Taiji (cyan), TianQin (magenta), and DECIGO (dark yellow).
}
\label{fig6}
\end{figure*}

Figure~\ref{fig6} shows that the characteristic strains of all four NSWD merger types are detectable by space-based GW observatories. 
Low-mass (LM) mergers produce narrowband, line-like signals similar to those of ultracompact X-ray binaries (UCXBs), while intermediate-mass (IM) and high-mass (HM) mergers generate broader-band signals with higher strain amplitudes.

For example, as shown in Fig.\,\ref{fig4}(c), the intermediate-mass NSWD merger with initial masses \( 1.40~\Msolar + 1.00~\Msolar \) 
evolves to \( 1.401~\Msolar + 0.383~\Msolar \) after unstable mass transfer, producing a strong GW strain 
(amplitude $h_{+}$: $\sim$0.4 $\times$ 10$^{-20}$) near \( f \sim 0.089~\mathrm{Hz} \) and a weaker 
one ($h_{+}$: $\sim$0.09 $\times$ 10$^{-20}$) near \( f \sim 0.0182~\mathrm{Hz} \) in a time period of $\sim$4 years. 

The coalescing binary with \( 1.35~\Msolar + 1.38~\Msolar \) follows a spectral amplitude scaling of \( \mathcal{A}(f) \propto f^{-7/6} \) for \( f \lesssim 0.7~\mathrm{Hz} \), consistent with the PhenomA waveform model for the in-spiral phase \citep{Ajith07}. At higher frequencies, deviations from this power law occur due to significant contributions from mass transfer to the orbital evolution, even though GW radiation remains dominant.

For a 4-year observation period, the SNRs of three NSWD mergers-\( 1.40~\Msolar + 0.07~\Msolar \), \( 1.35~\Msolar + 0.56~\Msolar \), and \( 1.40~\Msolar + 1.00~\Msolar \)-reach approximately 9/8/2/2, 119/183/132/368, and 74/89/163/25773 for LISA, Taiji, TianQin, and DECIGO, respectively. For the HM merger \( 1.35~\Msolar + 1.38~\Msolar \), observed over 0.158 years, the SNRs are 3/3/13/45132. These results suggest that LISA and Taiji are well-suited for detecting LM and IM mergers (G1, G2), TianQin is sensitive to IM mergers (G2), and DECIGO-like detectors are optimal for observing HM mergers (G4) with very high SNR.

Assuming a frequency change \( \Delta f \approx 0.42~\mathrm{Hz} \) over 0.158 years, the characteristic strain scales as \( h_{\rm c} \propto f^{1/2} \cdot f^{-7/6} = f^{-2/3} \), consistent with the slope shown in Fig.~\ref{fig6}.

Finally, we note that GW signals from accretion disks may appear as a low-level background. For demonstration, Fig.~\ref{fig6} includes the disk signal from the IM merger \( 1.40~\Msolar + 1.00~\Msolar \) in the conservative case, with amplitudes in the range \( \sim 10^{-22} \) to \( 10^{-21} \). 
The ISCO in this case is $\sim$12.5km, and the outer radius of accretion disk is estimated as $\lesssim1.08\times10^{4}$km.  
The maximum value of the radius comes from the orbital separation $1.66\times10^{4}$km subtracting 
the Roche lobe radius $0.58\times10^{4}$km of the WD. By comparison, the radii of fully developed disks for WD masses 
of 0.07$M_{\odot}$ and 1.2$M_{\odot}$ may be $\sim1.14\times10^{5}$km and $\sim7.22\times10^{3}$km.   
In coalescing systems, disk signals are generally absent due to the violent nature of the merger, which suppresses disk formation.

\section{Discussion}
\label{sec_discussion}

Using population synthesis methods, the Galactic merger rate of NSWD binaries has been theoretically estimated to lie in the range of \(10^{-6}\)-\(10^{-4}~\mathrm{yr}^{-1}\) \citep{Cooray04,Nelemans01b,Toonen2018}. 
The total number of such binaries in the Galaxy may reach up to \(2 \times 10^6\), of which approximately \(\mathcal{O}(10^2)\) are expected to be detectable by LISA \citep{Korol24}. 
Our calculations, based on similar methods, are broadly consistent with these estimates and further suggest merger rates of (1.4-34, 56-280, and 0.72-3.6) Myr\(^{-1}\) per Galactic mass (\(M_{\rm G} \approx 5 \times 10^{10}~\Msolar\), \citep{Cautun20}) for WD mass ranges of \(M_{\rm wd} < 0.07~\Msolar\), \(0.07 \leq M_{\rm wd} \leq 1.0~\Msolar\), and \(M_{\rm wd} > 1.0~\Msolar\), respectively, depending on the common envelope (CE) evolution parameters. 

\citet{Toonen2018} investigated the demographics of NSWD mergers using population synthesis approach.  
They adopted a variety of initial conditions and physical processes involved in the stellar evolution models 
of NSWD progenitors, including NS natal kicks, mass transfer in binaries, and common-envelope evolution. 
The upper limit of the total merger rates of NSWDs in the Galaxy may be $5\times10^{-4}\rm yr^{-1}$, slightly 
higher than that ($3.2\times10^{-4}\rm yr^{-1}$) in our computations. 
Our results in general agree with their findings that the majority of mergers have Carbon-Oxygen WD components ($\sim$61-89\%), 
but merger rates of NSWDs with He WDs in our models are several times higher than the rates of NSWDs with ONe WDs. 
The reasons for differences of the NSWD merger rates in these population synthesis studies will be studied in following work as many initial 
conditions and physical processes are involved. 

For high-mass (HM) NSWD mergers, we estimate the event rates of \(\sim 1,900\)-13,000 Myr\(^{-1}\) in nearby galaxy clusters and groups.
These estimates assume a total stellar mass of \(1.3 \times 10^{14}\) -- \(1.8 \times 10^{14}~\Msolar\) within a  \(\sim 50\) Mpc volume \citep{Dev2024,Madau2014} and 
a detection threshold of S/N = 5 by a DECIGO-like observatory. 
Variations in binary population parameters are expected to affect these rates by no more than one order of magnitude \citep{Nelemans01b,Yu10,Korol24}. 

\subsection{Equation of State}

Although the maximal NS mass and radius can be constrained by recent multimessenger data and 
latest theoretical calculations, a number of other calculations in the literature have also 
inferred the EoS, suggesting that the maximal NS mass can be up to $\sim$2.8$M_{\odot}$ and 
when NS mass goes down to 1$M_{\odot}$, NS radii vary between $\sim$9-15km (\citet{Ozel2016a} 
and references therein).  
\citet{Kalogera1996} found that the upper bound of NS mass can be up to 2.9$M_{\odot}$ 
by considering the EoS for NS matter as valid up to twice nuclear matter saturation density 
($\rho_{\rm nm}\sim 2.7 \times 10^{14} \rm g\,cm^{-3}$) 
and solving the Tolman-Oppenheimer-Volkoff (TOV) equations, while the lowest possible upper 
bound could be $\sim$2.2$M_{\odot}$ if increasing neutron matter up to 4$\rho_{\rm nm}$.

\begin{figure}
\hspace*{-0.5cm}
\centering
\includegraphics[width=9cm,clip,angle=0]{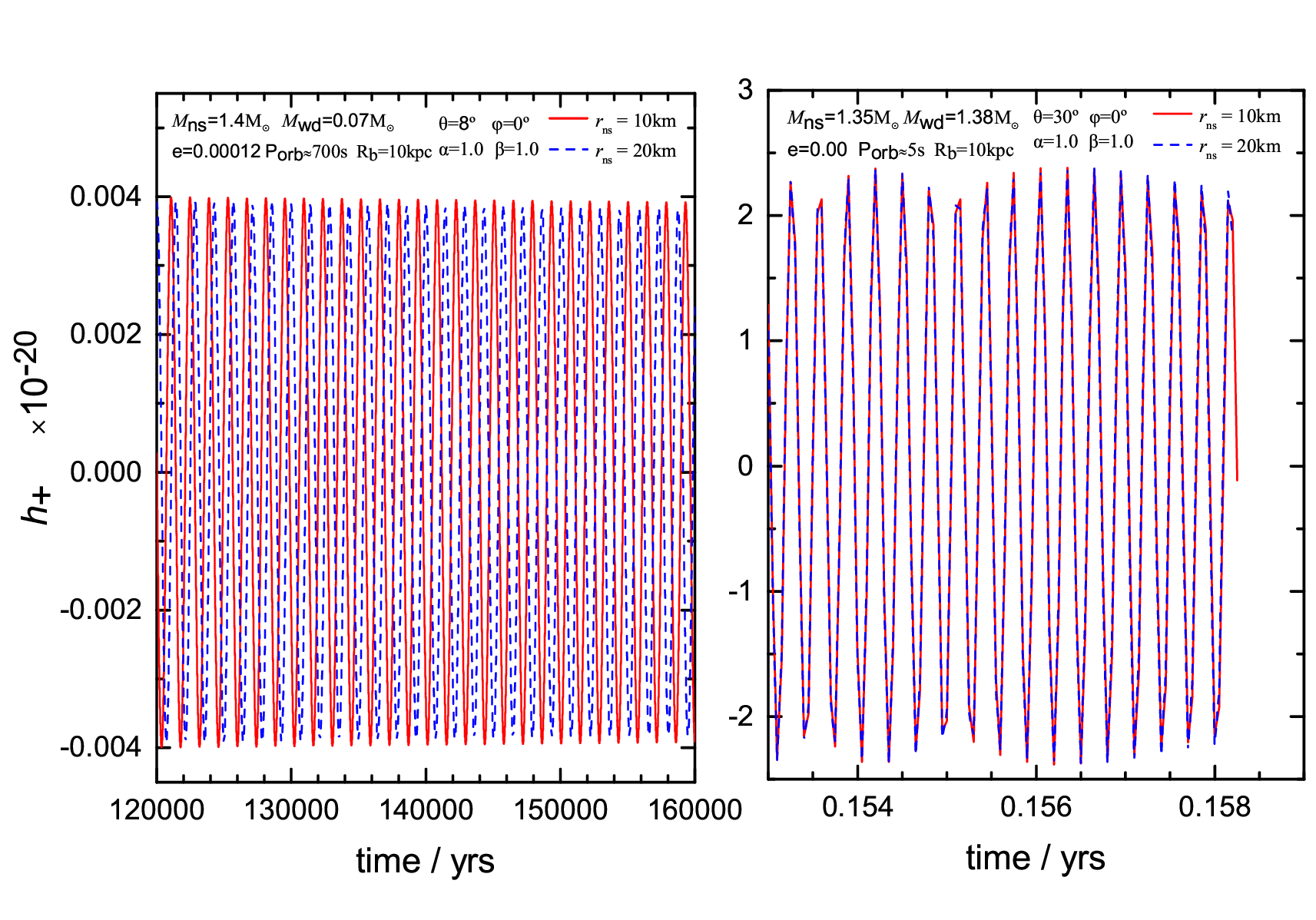}
\caption{
Comparison of the GW waveforms of merging NSWD binaries with NS radius of 
10km (red solid lines) and 20km (blue dashed lines). The initial parameters of the NSWD binaries 
in left panel and right panel are same with Fig.\,\ref{fig4}(a) and Fig.\,\ref{fig4}(d), 
respectively. 
Note that we do not plot the GW signals of post-mergers in the right panel.
}
\label{fig7}
\end{figure}

So we slightly loosen the NS radii from 10km to 20km in our fiducial model to investigate 
the variation of the GW waveforms from merging NSWDs. As shown in Figure\,\ref{fig7}, 
the waveforms can vary for the NSWDs with different masses and radii in their late-stage evolution. 
The differential of GW frequencies for the two low-mass binaries with the NS radii 10km and 20km 
in left panel is in the magnitude of $\sim10^{-9}\rm Hz\,yr^{-1}$, while the maximal frequency differential 
for the two high-mass binaries in right panel is in the magnitude of $\sim10^{-4}\rm Hz\,day^{-1}$. 
The GW frequency differential arises from the change of gravitational potential at the surface of neutron star 
(cf. Eq.\,(23) in \citet{Yu21}). Measuring the GW amplitude and frequency by space-based detectors with 
sufficient sensitivity might be a manner to constrain the NS mass and radius, and further infer the EoS.
Note that we do not plot the GW signals of post-mergers in the right panel of Fig.\,\ref{fig7}.

\subsection{Multi-messenger implications}

Intermediate- and high-mass NSWD mergers may produce observable explosive electromagnetic (EM) transients. Intermittent RLOF in NSWD binaries has been proposed as a source of repeating fast radio bursts (FRBs) \citep{Gu16,Zhong20}. Our results support this hypothesis and further suggest that coalescing NSWDs could contribute to non-repeating FRBs. The isotropic luminosity from accretion can reach
\[
L \sim \frac{G M_{\rm ns} \dot{M}_{\rm wd}}{r_{\rm ns}} \approx 1.7 \times 10^{46}~\mathrm{erg~s^{-1}},
\]
assuming \(M_{\rm ns} = 2~\Msolar\), \(r_{\rm ns} = 10~\mathrm{km}\), and \(\dot{M}_{\rm wd} = 1~\Msolar~\mathrm{yr^{-1}}\). 
This is comparable to the upper limit of isotropic-equivalent peak luminosities observed in FRBs \citep{Zhang20}. 

The total NSWD merger rate may lie in the range \((0.04-4) \times 10^6~\mathrm{yr^{-1}~Gpc^{-3}}\), while radio observations constrain FRB rates above \(10^{42}~\mathrm{erg~s^{-1}}\) to \((1.1-9.2) \times 10^4~\mathrm{yr^{-1}~Gpc^{-3}}\) \citep{Zhang20,Luo20}. 
The observable upper limit of the FRB rates due to NSWD mergers in our model can be 
estimated as $f(\sigma)\times$\(0.92 \times 10^4~\mathrm{yr^{-1}~Gpc^{-3}}\). 
The average fraction, $f(\sigma)$, of the solid angle swept out by beams of angular 
radius $\sigma$ (in radians) centered on the north and south magnetic poles is expressed as 
\citep{Emmering89} 
\begin{equation}
f(\sigma)=(1-\cos\sigma)+(\frac{\pi}{2}-\sigma)\sin\sigma.
\end{equation}
Adopting $20^{\circ}<\sigma<50^{\circ}$ (a typical value for millisecond pulsars), 
the beaming fraction is $0.5<f(\sigma)<0.9$ \citep{Kramer98}.  
This suggests that NSWD mergers could represent a significant fraction of the FRB population. 
Their spatial distribution and contribution could be further constrained through joint radio and GW observations in the millihertz to decihertz band.

The continuous gravitational waves generated by r-mode fluid oscillation with the restoring force 
being the Coriolis force in rotating neutron stars have been studied and searched in observation data of the LVK 
\citep{Dunn2025,Covas2025,Tenorio2025,Riles2023,Wette2023,Abbott2022,Haskell2015b}.
The power spectral density $S(f)$ of r-mode induced GW emission 
for the neutron stars with low accretion rate ($10^{-8}M_{\odot}\rm yr^{-1}$) 
is $S(f)^{1/2}\lesssim1.3\times10^{-21}\rm Hz^{-1/2}$ in $\sim$10-1000Hz, 
which is hardly to be detected by the LVK collaboration \citep{Dong2025}. 
It would be worth studying the GW amplitude from neutron stars with high accretion rate, 
since the amplitude is proportional to the accretion rate 
and the GW signals could appear in high-frequency bands of LISA-type detectors.  

\subsection{Electromagnetic counterparts and nucleosynthesis}

NSWD mergers may also produce sub-luminous supernova-like transients, as indicated by smoothed particle hydrodynamics and hydrodynamical-thermonuclear simulations \citep{Zenati19,Zenati20,Metzger12}. 
For a \(1.4 + 0.6~\Msolar\) NS-CO WD binary, more than \(0.3~\Msolar\) of material may be ejected at velocities of \(1$-$5 \times 10^4~\mathrm{km~s^{-1}}\), with up to \(0.006~\Msolar\) of \(^{56}\mathrm{Ni}\) synthesised and bolometric luminosities reaching \(\sim 6 \times 10^{40}~\mathrm{erg~s^{-1}}\). 
Shock heating from late-time outflows could increase the peak luminosity to \(\sim 10^{43}~\mathrm{erg~s^{-1}}\) \citep{Margalit16,Fernandez19}. 

\citet{Zenati20} estimated that explosions resulting from NSWD mergers could produce observable transients 
at rates of up to $\sim$10–70 yr$^{-1}$, assuming the predicted efficiency for the Large Synoptic Survey Telescope 
in R and I bands. 
NSWD merger events may correspond to certain supernova-like phenomena that exhibit not only radio 
and X-ray emissions but also UV, optical, and near-infrared signatures, such as SN 2019ehk \citep{Jacobson-Galan20}, 
SN 2019wxt \citep{Shivkumar23}, and SN 2022oqm \citep{Yadavalli24}. 
Thus, observations in both electromagnetic and gravitational waves, combined with simulations of NSWD systems 
across a range of masses, may shed light on the roles of mass transfer, nuclear burning, 
and magnetic fields in shaping chemical abundance evolution. 

\subsection{Accretion disks and GW signatures}

The GW contribution from accretion disks is closely tied to mass outflows. 
Using the Eddington limit to constrain the accretion rate, we find that in a \(1.4 + 0.1~\Msolar\) NSWD binary, more than 50\% of the WD's lost mass can be accreted by the NS. 
In contrast, for coalescing systems, the accretion efficiency \(\alpha\) drops sharply to \(\sim 0.1\)-0.01\%. 
One-dimensional steady-state models suggest that at high accretion rates (\(10^{-4}$-$10^{-1}~\Msolar~\mathrm{s^{-1}}\)), most of the disk becomes radiatively inefficient and prone to outflows driven by viscous dissipation and nuclear burning. 
For plausible wind properties, 50-80\% of the WD mass may be unbound \citep{Metzger12,Margalit16}. 
Our results indicate that only under conservative mass transfer conditions can a massive disk form, consistent with recent simulations \citep{Kaltenborn23,Paschalidis11}. 
However, the compact core of a massive WD may survive intense mass transfer, producing only weak GW signals.

\subsection{Comparison with previous studies}
\citet{Paschalidis2009} investigated NSWD binaries in close 
binary separations, and found that when mass transfer 
from the WD onto the NS across the inner Lagrange point occurs, 
it can either be stable or unstable. This result is similar to the 
studies of \citet{Verbunt88} and \citet{Toonen2018}. 
The unstable mass transfer scenario is considered as the tidal disruption 
of the WD by the NS. Similar research is presented in \citet{Moran-Fraile2024}. 

In our study, we assume that the quasiequilibrium sequences 
describing the contact NSWD binaries are not completely broken if the orbit of 
a binary can significantly expand, and a less massive compact core of white dwarf may 
survive. A significant fraction of mass loss from WDs may form common envelope 
around NSWD binaries. This scenario happens for the IM mergers, while for the HM mergers, 
disruption of WDs and coalescence take place. This is why we see two main peaks in the evolution 
track of 1.40 $M_{\odot}$ + 1.00 $M_{\odot}$ in Fig.\,\ref{fig6} (Section\,\ref{sec_snr}),  
but only see one main peak in Fig.8 in \citet{Moran-Fraile2024}.

\subsection{Outlook}

Detecting NSWD mergers will provide valuable insights into their progenitor populations \citep{Chen20,Korol24,Lu25}, test models of compact binary evolution \citep{Wang21,Badenes09}, and potentially allow direct measurement of mass transfer rates \citep{Lyu24}. 
A comprehensive understanding of NSWD evolution from in-spiral to merger across the full parameter space will be essential for developing accurate GW waveform templates, which are critical for interpreting future detections.

\section{Conclusions}
\label{sec_conclusions}
We have investigated the evolution of NSWD binaries during the mass transfer phase, and 
have computed their orbital evolution as a function of four key parameters: neutron star mass, white dwarf mass, orbital eccentricity, and accretion fraction. 

Our results reveal well-defined boundaries in the NS-WD mass-mass parameter space.
Systems with white dwarf masses exceeding these thresholds experience rapid orbital decay, leading to prompt coalescence. 
The position of these boundaries depends sensitively on system parameters, indicating that Roche-lobe-filling NSWD binaries can evolve along multiple distinct pathways—a phenomenon we refer to as  branched or polymorphic evolution.

These systems emit strong and varied gravitational wave signals, many of which lie within the detection band of space-based GW observatories. 
The evolving waveform morphology encodes detailed information about the binary configuration, including potential contributions from an accretion disk. 
Our models provide critical waveform templates for the identification and real-time classification of merging NSWD binaries in GW data.

\section*{Acknowledgments}
We thank the referee, Yossef Zenati, for the insight and helpful comments and suggestions. 
This work was supported by the National Key Research and Development Program of China (No. 2020YFC2201400 and No. 2021YFC2203003), 
the National Natural Science Foundation of China (Grant Nos 11673031, 11690024), and CAS Project for Young Scientists in Basic Research, Grant No. YSBR-063. 
This work was also supported by the science and technology innovation Program of Hunan Province (No. 2024JC0001). 
Research at the Armagh Observatory and Planetarium is grant-aided by the N. Ireland Department for Communities. 
This work made use of the data from FAST (Five-hundred-meter Aperture Spherical radio Telescope). 
FAST is a Chinese national mega-science facility, operated by National Astronomical Observatories, Chinese Academy of Sciences. 
Starting from an original manuscript, Microsoft CoPilot was used to improve the language of this paper. 
This was carried out section by section in response to the following request: "Please redraft the following, 
correcting and simplifying the English where appropriate, bearing in mind it is for publication in a scientific journal."

\section*{DATA AVAILABILITY}
Some or all data, models, or code generated or used during the study
are available from the corresponding author by request.

\bibliographystyle{mnras}
\bibliography{gwrlof}

\label{lastpage}

\end{document}